%% file: merge.tex
\documentclass[
twocolumn,aps,prl,showpacs,superscriptaddress
]{revtex4}

\usepackage{subfiles}

\usepackage{graphicx,color,hyperref}

\usepackage{amsmath}
\usepackage{amsfonts}
\usepackage{amssymb}

\usepackage{braket}

\newcommand{\tr}{\mathrm{tr}}

\usepackage[normalem]{ulem} 

\begin{document}

\title{
Quantum ergodicity in the many-body localization problem
}

\author{
Felipe Monteiro
}
\affiliation{Centro Brasileiro de Pesquisas F\'isicas, Rua Xavier Sigaud 150, 22290-180, Rio de Janeiro, Brazil
}

\author{Masaki Tezuka}
\affiliation{Department of Physics, Kyoto University, Kyoto 606-8502, Japan
}

\author{Alexander Altland}
\affiliation{Institut f\"ur Theoretische Physik, Universit\"at zu K\"oln, Z\"ulpicher Str. 77, 50937 Cologne, Germany
}

\author{David A. Huse}
\affiliation{Department of Physics, Princeton University, Princeton, NJ 08544, USA}

\author{
T.~Micklitz
}
\affiliation{Centro Brasileiro de Pesquisas F\'isicas, Rua Xavier Sigaud 150, 22290-180, Rio de Janeiro, Brazil
}

\subfile{letter2-rev.tex}

\clearpage

\setcounter{equation}{0}
\renewcommand{\theequation}{S\arabic{equation}}

\subfile{suppMat.tex}

\end{document}

%% file: letter2-rev.tex
\date{\today}


\begin{abstract}

We generalize Page's result on the entanglement entropy of random pure states   
 to the many-body eigenstates of realistic disordered many-body systems subject to long range interactions. This extension leads to two principal conclusions: first, for increasing disorder the ``shells'' of constant energy supporting a system's eigenstates
 fill only a fraction of its full Fock space and are subject to intrinsic correlations absent in synthetic high-dimensional random lattice systems. Second, in all regimes preceding the many-body localization transition individual eigenstates are thermally distributed over these shells. These results, corroborated by comparison to exact diagonalization for an SYK model,  are  at variance with the concept of ``non-ergodic extended states'' in many-body systems discussed in the recent literature. 
\end{abstract}

\maketitle

{\it Introduction:---} Complex quantum systems exposed to 
external disorder may enter a phase of strong localization. About two decades after
the prediction of many-body localization
(MBL)~\cite{AKGL,BaskoAleinerAltshuler,Mirlin},  there is still no strong consensus
about the stability of the MBL phase and/or the  possible presence of an intermediate phase
between  MBL  and the thermal phase. One class of models where these
questions can be explored with  more analytic control is confined many-body systems
  with long-range interactions.  
Under these conditions, the interaction operator couples all single-particle states, which facilitates the analysis.  At the same time, the  Hilbert space dimension is still exponentially large in the particle number, which leads to   rich
physics relevant to systems such as  chaotic many body quantum devices~\cite{AltshulerGefenKamenevLevitov97,Silvestrov97,Silvestrov98,GornyMirlinPolyakov16,GornyMirlinPolyakovBurin17}, 
small sized optical lattices~\cite{RubioAbadal19,Choi16,Schreiber15}, or qubit arrays~\cite{kai18, roushan17}.

In recent years, the complex structure of many-body quantum \emph{states} in MBL has
become a focus of intensive research. Unlike with single particle problems, where
extended wave functions   uniformly cover real space, increasing the disorder in a
phase of extended many body states $|\psi\rangle$ leads to a diminished wave function
support in Fock space. This phenomenon, which  shows, e.g., in a suppression of  wave
function moments (WFM) $|\langle n|\psi\rangle|^{2q}$ in an occupation number basis,
$|n\rangle$, has led to the proposal of a   phase of ``\emph{nonergodic} extended
states''~\cite{NonE_Ex,RPmodel1,Kravtsov1,Kravtsov2} intermediate between the phases
of absent and strong localization. An alternative scenario is that for each
realization of the disorder only a subset of states $\{|n\rangle\} $ have finite
overlap with the eigenstates of energy $E$, and in this way define a quantum
\emph{energy shell} in Fock space. A uniform (thermal) distribution of the exact
eigenstates on this shell would then be the defining criterion for maintained quantum
\emph{ergodicity} on the delocalized side of the MBL transition.

At this stage, there is mounting evidence in favor of the second scenario~\cite{Mirlin1,TikhonovMirlin19_2,TikhonovMirlinRev21,TikhonovMirlinPRB21}. However,
in order to firmly characterize the physics of a globally realized many body ergodic
quantum phase, two questions need to be addressed: how can the energy shell be
described in quantitative terms? And what  is the distribution of quantum states on that
shell? As indicated above, wave function statistics can provide at least part of an
answer to the first question. In this Letter, we focus on the equally important second
part of the problem and demonstrate that the key to its solution lies in  concepts of
quantum information. Specifically, we will compute pure state entanglement entropies
(EE) under a relatively mild set of assumptions. Within this framework we find that
to zeroth order wave functions remain thermally distributed over the shell. This
establishes a microcanonical distribution, in agreement with the second scenario --- maintained ergodicity in all
regimes prior to the transition. In addition, the EE contains
sub-leading terms which reflect the characteristic way in which the energy shell is
interlaced into Fock space. These contributions sharply distinguish the energy shells
of genuine many body systems from those of phenomenological high dimensional models
such as the random energy model (REM), or sparse random states~\cite{TomasiKhatmovich2020}. In this way the
combined analysis of WFMs and EEs becomes a sensitive probe into
the complex manifestation of wave function ergodicity in many particle systems.

  \emph{Pure state entanglement
entropies: ---}  For a  pure state, $\rho = |\psi\rangle \langle \psi|$, the entanglement entropy relative to a partitioning  $\mathcal
F=\mathcal F_A
\otimes \mathcal F_B$ of Fock space is defined as the von Neumann entropy,
$S_A=-\mathrm{tr}_A(\rho_A \ln \rho_A)$ of the reduced density matrix $\rho_A
=\mathrm{tr}_B(|\psi \rangle \langle \psi |)$ . The entanglement entropies of pure maximally random states were calculated in the classic Ref.~\cite{Page93}. More recent work~\cite{HaqueKhaymovich2020} emphasizes the utility of the concept in the context of random matrix models serving as proxies of high-dimensional localizing systems~\cite{RPmodel1}.  In these systems, quantum interference shows in a contribution to the entanglement entropy proportional to the ratio of subsystem Fock-space dimensions. A main finding of the present work is that   energy-shell correlations distinguishing microscopic systems from random matrix models open a second channel of quantum information and exponentially enhance the suppression of the entanglement below its thermal value.  In this way, the entanglement sharply distinguishes between genuine many-body wave functions and wave functions on generic high-dimensional random lattices.

In the rest of this Letter, we will compute the entanglement entropy of pure states
prior to the onset of strong localization  under a minimal set of assumptions. We
will compare our results to the entropies obtained for phenomenological models and to numerical data  obtained for a Majorana SYK model.

\emph{Energy shell: ---} We begin with a qualitative discussion of the Fock space energy shell. Consider a
many-body Hamiltonian $\hat H=\hat H_2 + \hat H_4$, where $\hat H_4$ is an interaction operator and $\hat H_2$  a one-body operator
defined by a single particle spectrum $\{m_i\}$, $i=1,\dots, N$
distributed over a range $\delta$. Working in the eigenbasis of $\hat H_2$, Fock
space is spanned by the $D\equiv 2^N$ occupation number states $n=(n_1,\dots,n_N)$,
$n_i=0,1$ for spinless fermions. We interpret these states as sites of a hypercubic
lattice, carrying local potentials  $v_n=\sum (2n_i-1) m_i$ with r.m.s. value
$\Delta_2\equiv N^{1/2}\delta$. Individual  states $n$ are connected to a
polynomially large number $N^\alpha$ of `nearest neighbors' $m$ by the 
interaction $\hat H_4$. For interaction matrix elements  $t_{nm}\sim g
N^{-\beta/2}$, the r.m.s. eigenvalue of $\hat H_4$ scales as $\Delta_4
\sim gN^{(\alpha-\beta)/2}$, with $g$ an $N$-independent coupling energy for the interaction.  These interactions change only an order-one number of occupation numbers, so $|v_n-v_m|$ is of order $\delta$ and thus for large $N$ much smaller than the `bandwidth' $\Delta_2$ of $\hat H_2$.

In the competition of the operators $\hat H_2$ and $\hat H_4$,  states $n$ may
hybridize with states $m$ via the coupling $t_{nm}$. 
When the eigenstates of $\hat H$ are delocalized in Fock space, 
this hybridization gives the local spectral density 
\begin{align}
    \label{LocalDosLorentzian}
    \nu_n(E)\equiv -\frac{1}{\pi}\mathrm{Im} \langle n|(E^+-\hat H)^{-1}|n\rangle, 
\end{align}
a linewidth $\kappa=\kappa(v_n,\delta,g)$ which must be self-consistently
determined~\cite{fn1}. The solution of Eq.~\eqref{LocalDosLorentzian} for a given realization of the disorder contains the essential information on the distribution of the energy shell in Fock space. Specifically, for generic values of the energy $E$ (we set
$E=0$ for concreteness), the strength of the disorder, $\delta$, defines four regimes of different shell structure:

\noindent\textit{I:} $\delta\ll N^{-1/2}\Delta_4$:  the characteristic disorder band width
$\delta N^{1/2}= \Delta_2\ll\Delta_4$ is perturbatively small. In this regime, the
spectral density, $\nu_n\equiv \nu$ is approximately constant over energy scales
$\sim\Delta_2$. 

\noindent\textit{II:} $N^{-1/2}\Delta_4\ll\delta\ll\Delta_4$: the  bandwidth of
$\hat H_2$ 
exceeds that of the interaction $\hat H_4$, but nearest neighbors remain
energetically close  $|v_n - v_m|\sim \delta \ll \Delta_4$. In this regime,  $\kappa=\Delta_4$,
indicating that the full interaction Hamiltonian enters the hybridization of neighboring sites.

\noindent\textit{III:} $\Delta_4\ll\delta \ll \delta_c$: 
only a fraction $\sim (\Delta_4/\delta)^2$ of nearest neighbors
remain in resonance, and the broadening is reduced to $\kappa\sim \Delta^2_4/\delta$. 

\noindent\textit{IV:} The threshold to localization, $\delta_c$, is reached when less than one of the $\sim N^\alpha$ neighbors of characteristic energy separation $\delta$ falls into the broadened energy window. 
Up to corrections logarithmic in $N$ (and neglecting potential modifications due to Fock space loop amplitudes) this leads to the estimate $\delta_c \sim N^{\alpha/2}\Delta_4$ for the boundary to the strong localization regime.

The energy shell in the delocalized regimes II and III is an extended cluster of 
 resonant sites embedded in Fock space. It owes its structure to the  competition between the large number  $\mathcal{O}(N^\alpha)$ of nearest neighbor matrix elements and the
 detuning of  statistically correlated nearest neighbor energies, $v_n,v_m$.
 In regime II, only a polynomially (in $N$) small fraction 
 $\kappa/\Delta_2\stackrel{II}\sim \Delta_4/(\delta N^{1/2})$ of Fock space sites lie in the resonant window 
 defining the energy shell, and in III this fraction is further reduced to
$\stackrel{III}\sim \Delta_4^2/(\delta^2 N^{1/2})$, before the shell fragments at the boundary to regime IV.  

We also note that \emph{if} a site, $n$, lies on the shell, the probability that its
neighboring sites of energy $v_m = v_n\pm
 \mathcal{O}(\delta)$ are likewise on-shell is parametrically enhanced 
 compared to
 that of generic sites with energy $v_n\pm \mathcal{O}(\Delta_2)$. It is this principle which gives the energy shell of many-body systems a high degree of internal correlations (absent in phenomenological lattice models with statistically independent on-site randomness)~\cite{fn2}. 
  What physical quantities are
 sensitive to these correlations? And how do quantum states spread over the shell structure? As we are going to discuss next,
 the pure state entanglement entropy, $S_A$, contains the answer to these questions. 

\begin{figure}[t!]
\centering
\includegraphics[width=8.5cm]{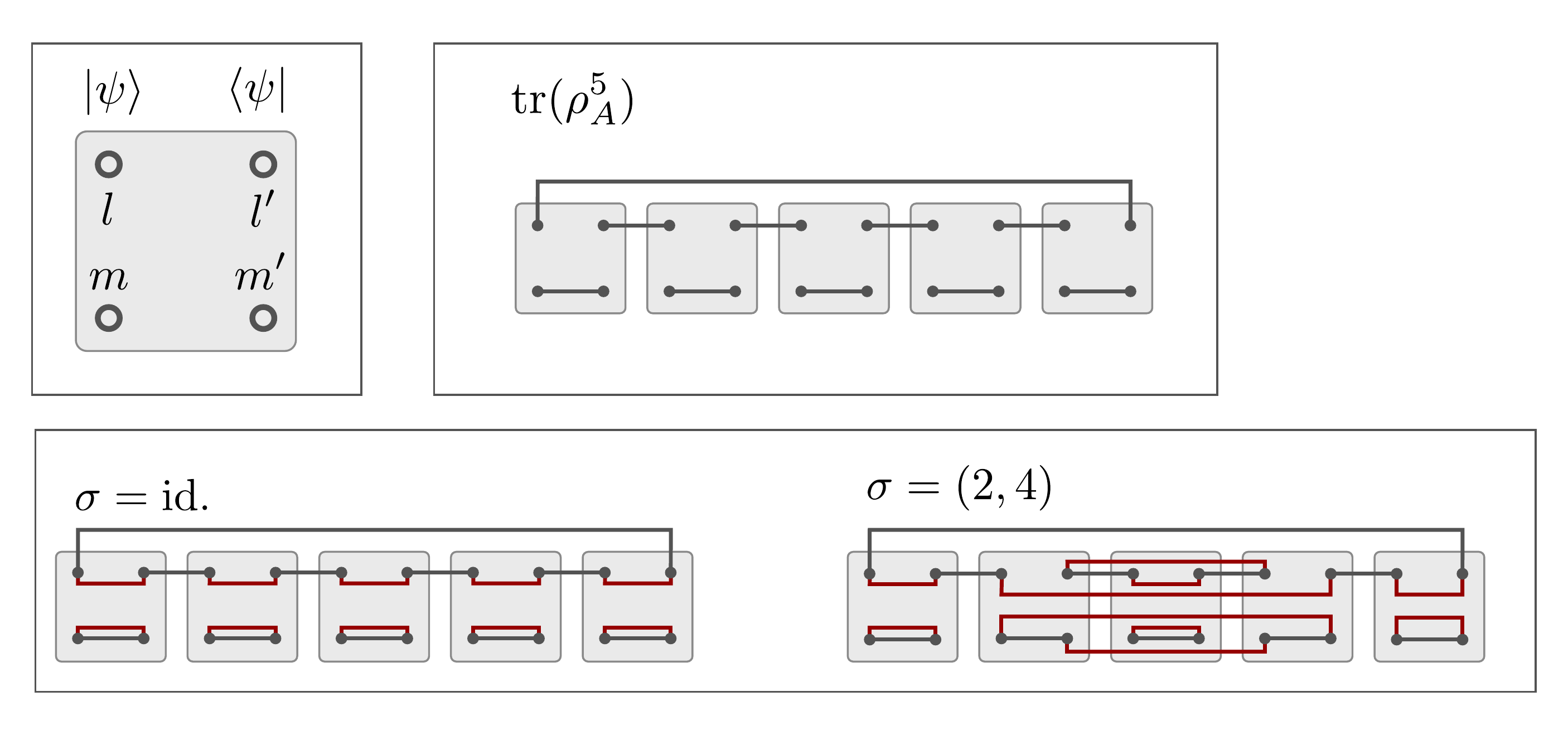}
\caption{\label{fig2}
Top left: graphic representation of the tensor amplitude $\psi_{lm}\bar \psi_{l'm'}$. Top right: contraction of indices defining $\tr(\rho_A^5)$. Bottom: averaging enforces pairwise equality of indices $n,n'$ in tensor products $\langle \ldots \psi_{n}\ldots \bar \psi_{n'}\ldots\rangle$, as  indicated by red lines. Left: identity pairing of indices within the five factors $\langle \mathrm{tr}_A(\rho_A \rho_A\rho_A\rho_A\rho_A)\rangle$. Right: pairing of indices of the second and fourth factor.
}
\end{figure}

\emph{Entanglement entropy: ---} Consider a Fock space (outer product) partitioning defined by
 $n=(l,m)$ where the $N_A$-bit vector $l$ labels the states of subsystem $A$ and $m$ those of $B$ with 
 $N_B=N-N_A\gg N_A$. We are interested in the disorder
 averaged moments $M_r\equiv
\langle\mathrm{tr}_A(\rho_A^r)\rangle$, and the entanglement entropy
$S_A=-\partial_{r} M_r|_{r=1}$ of the reduced density matrix, $\rho_A
=\mathrm{tr}_B(|\psi \rangle \langle \psi |)$, defined by a realization-specific
zero-energy eigenstate $\hat H |\psi\rangle =0$.  The bookkeeping of index configurations entering the
moments  $\mathrm{tr}_A(\rho_A^r)=\psi_{l^1 m^1}\bar \psi_{l^2 m^1}
\psi_{l^2 m^2} \dots \psi_{l^rm^r }\bar \psi_{l^1 m^r}$ is conveniently done in a
tensor network representation as in Fig.~\ref{fig2}. Introducing a multi-index
$\mathcal{N}\equiv (n^1,\dots,n^r)$, and analogously for $\mathcal{N}_{A,B}$, the
figure indicates how the index-data $\mathcal{N}$ and $\mathcal{M}$ carried by $\psi$ and  $\bar \psi$
 is constrained by the summation as
$\mathcal{M}_B^i=\mathcal{N}_B^i$ and $\mathcal{M}_A^i=\mathcal{N}_A^{\tau i}$, where
$\tau i = (i+1)\mathrm{mod}(r)$. A further constraint, indicated by red lines in the
bottom part of the figure, arises from the random phase cancellation under averaging,
which in the present notation requires $\mathcal{N}^i\equiv
\mathcal{M}^{\sigma i}$, for \emph{some} permutation $\sigma$. (The figure
illustrates this for the identity, $\sigma =\mathrm{id}.$, and the transposition
$\sigma= (2,4)$.) Combining the two constraints, we obtain the representation $
M_r=\sum_\sigma \sum_\mathcal{N} \prod_i \langle |\psi_{n^i}|^2\rangle
\delta_{\mathcal{N}_A,\sigma \circ \tau \mathcal{N}_A} \delta_{\mathcal{N}_B,\sigma
\mathcal{N}_B}$. This expression is universal in that it does not require  assumptions other than the random phase cancellation. 
In a less innocent final
step we establish contact to the previously discussed local density of states, $\nu_n$, and compare the two
representations $D\nu \equiv \sum_\alpha \delta(E-E_\alpha)=\sum_{n,\alpha} |\psi_{\alpha,n}|^2\delta(E-E_\alpha)=\sum_n
\nu_n$ to identify $|\psi_n|^2
=\frac{\nu_n}{D \nu}$. In other words, we identify the moduli $|\psi_{n}|^2$ of a fixed
eigenstate $\psi=\psi_\alpha$ with the realization specific
local density of states, $\nu_n$, at  $E=E_\alpha$. For the legitimacy of this replacement for single particle random systems see Ref.~\cite{Prigodin}, and for the SYK model  the Supplemental Material and Ref. \cite{SYK4+2}. With this substitution, we obtain the  representation 
\begin{align}
\label{moments_gen}
M_r
=
\sum_{\sigma}
\sum_{\cal N}\prod_{i=1}^r \lambda_{n_i}\,
\delta_{{\cal N}_A, (\sigma\circ \tau){\cal N}_A}
\delta_{{\cal N}_B, \sigma {\cal N}_B},
 \end{align}
with $\lambda_n\equiv \frac{\nu_n}{D \nu}$. This expression describes two complementary perspectives  of quantum states in Fock space: their support on a random energy shell defined by the coefficients $\lambda_n\sim \nu_n$, and random phase cancellations implicit in the combinatorial structure. In the following, we discuss the manifestations of these principles in the above regimes I-IV.

\emph{Regime I, maximally random states: ---} Here, 
 wave functions are uniformly distributed, $\nu_n =\nu$, and the evaluation of
 Eq.~\eqref{moments_gen} reduces to a combinatorial problem. The latter has has been
 addressed in the string theory literature~\cite{Stanford20,Liu20}
 (where high-dimensional pure random states are considered as proxies for black hole
 micro states.) 
 Inspection of the formula shows that increasing permutation complexity needs to be
 paid for in summation factors $D_B$. 
Keeping only the leading term,
$\sigma=\mathrm{id.}$, and the next leading single
transpositions $\sigma=(ij)$, we obtain $M_r\approx
D_A^{1-r}+\binom{r}{2}D_A^{2-r}D_B^{-1}$, and the subsequent differentiation in $r$ yields Page's result~\cite{Page93}
 \begin{align}
\label{S_page}
S_A-
S_{\mathrm{th}}= - \frac{D_A}{2D_B},\qquad S_{\mathrm{th}}=\ln D_A.
\end{align}
Interestingly, higher order terms in the $D_A/D_B$-expansion vanish 
 in the replica limit~\cite{Page93,proof1,proof2,Stanford20,Liu20}, and Eq.~\ref{S_page} is
exact for arbitrary $N_A\le N_B$, up to corrections small in $1/D$. (The case $N_A\ge N_B$ follows from exchange $A\leftrightarrow B$.) The result states that to leading order the entropy of the subsystem is that  of a maximally random (`thermal')
state, $S_\mathrm{th}$. The residual term results from wave function  interference  across system boundaries. Reflecting a common signature of
`interference contributions' to physical observables,  it is suppressed
by a factor proportional to the Hilbert space
dimension.

\emph{Regime II \& III, energy shell entanglement: ---} The energy
shell now is structured and correlations in the local densities, $\{\nu_n\}$, lead to a
much stronger correction to  the thermal entropy. Since these contributions come from
the identity permutation (do not involve wave function interference), we ignore 
for the moment
$\sigma\not=\mathrm{id.}$, reducing Eq.~\eqref{moments_gen} to $M_r \simeq \sum_l
\lambda_{A,l}^r$ with $\lambda_A\equiv \mathrm{tr}_B(\lambda)$. This expression
suggests an interpretation of the unit normalized density  $\{\lambda_n\}$
as a \emph{spectral measure}, $\sum_n \lambda_n=1$, $\lambda_n\ge 0$, and of
$\lambda_A$ as the reduced density of system $A$. With this identification,
the entropy,
\begin{align}
    S_A \approx S_\rho \equiv \mathrm{tr}_A (\lambda_{A} \ln(\lambda_{A}))
\end{align}
becomes the information entropy of that measure.

This is as far as the model-independent analysis goes. Further progress is contingent on two assumptions, which we believe should be satisfied for a wide class of systems in their regimes II and III:  
First, the exponentially  large number of sites entering the computation of the spectral measure justifies a self averaging assumption, 
\begin{align}
\label{eq:GaussianAverage}
    &\sum_{n_X} F(v_{n_X})\approx D_X\langle F(v_X)\rangle_{X}\equiv \cr
    &\qquad \equiv \frac{D_X}{\sqrt{2\pi} \Delta_X}\int dv_X\, \exp\Big(-\frac{v_X^2}{2\Delta_X^2}\Big)F(v_X),
\end{align}
where $X=A,B,AB$ stands for the two subsystems, or the full space, respectively,
$D_X$ are the respective Hilbert space dimensions, and $\Delta_X=\delta \sqrt{N_X}$. In
other words, we replace the sum over site energies by an average over a single variable whose Gaussian distribution follows from the central limit theorem. 
Second, when integrated against the distribution of subsystem energies $v_B$,   
the local DoS at zero energy $E\simeq0$ acts as a smeared $\delta$-function, setting the additive energy $v= v_A+v_B\simeq 0$, and effectively smoothening the distribution $\lambda_{A,l}$. 
Since $\kappa\ll \Delta_2\sim\Delta_B $, the detailed value of the width of the shell, $\kappa$, is of no significance in this construction.

Under these assumptions, straightforward computations detailed in the supplementary material yields, e.g., the density of states as $D\nu=\sum_{AB} \nu_n\approx D \langle  \delta_\kappa(v)\rangle_{AB}=\frac{D}{\sqrt{2\pi N}\delta }$. Applied to the computation of the moments Eq.~\eqref{moments_gen}, the averaging procedure obtains the entanglement entropy as~\cite{SuppMat} 
\begin{align}
    \label{EERegimeIII}
    S_A-S_{\mathrm{th}}
    &= 
    -\frac{1}{2}\ln\left(\frac{N}{N_B}\right)+\frac{1}{2}\frac{N_A}{N}
    -\sqrt{\frac{N}{2N_A}}\frac{D_A}{2D_B}.
\end{align}
 A number of  comments on Eq.~\eqref{EERegimeIII}: Provided the above assumptions on the  spectral measure hold, the result has the same level of rigor as Page's formula Eq.~\eqref{S_page}. The main difference is that  (for  small
subsystems, $N_A\ll N$) the information entropy $S_A-S_{\mathrm{th}}
  \approx -\frac{1}{4}\left(\frac{N_A}{N}\right)^2$ 
is exponentially enhanced compared to the correction in Eq.~\eqref{S_page}. Also note
that there is no dependence on the disorder strength (see supplemental material for more details).   

\emph{Comparison to phenomenological models: ---} The entanglement
entropy~\eqref{EERegimeIII} is a universal signature of \emph{correlations} (but not
the volume) of the energy shell. Conversely, the WFMs, $|\psi_n|^{2q}$, describe the
shrinking of the shell volume (but not its correlations). To see that these are
independent pieces of information, it is instructive to  compare to  the random
energy model (REM)~\cite{REM}, a phenomenological model  replacing the one-body
randomness by a set of statistically independent Fock state potentials $\{v_n\}$. For
increasing $\delta$, the WFMs diminish as  in microscopic models~\cite{SYKRE}. However,
we have verified that the EE of REM states coincides with Page's Eq.~\eqref{S_page}.
The same result is obtained for sparse random states~\cite{TomasiKhatmovich2020}, as
even more phenomenological proxies of many body states. What is the origin of the
difference to Eq.~\eqref{EERegimeIII}? A genuine many-body model describes many
``bodies'', representing the microscopic degrees of freedom.  The Fock space is an
outer product over the single body spaces, and the Hamiltonian  contains only
operators coupling $\mathcal{O}(1)$ of these degrees of freedom.  In this
sense the REM is {\it not} a many-body model, since its nonlocal energy operator acts
on the  products of all (or most) degrees of freedom simultaneously. Specifically, it lacks the principle of energy subsystem additivity $E=E_A + E_B$, required by Eq.~\eqref{EERegimeIII}. In this way, the entanglement
entropy  becomes a sensitive 
indicator of whether quantum states are genuine many body states or of different
origin.

 \emph{Regime boundaries: ---} Upon approaching the boundary to the trivially ergodic regime I, the second condition gets compromised, i.e. the
width $\kappa$ of individual states ceases to be small compared to the statistical
fluctuations 
$\sim \Delta_B$. Leaving a detailed analysis of the crossover region to future work, our numerics below shows a collapse of Eq.\eqref{EERegimeIII} to 
Eq.~\eqref{S_page} upon crossing the regime boundary. 
In the opposite MBL regime IV,  eigenstates are concentrated on a small number 
${\cal O}(1)$ of isolated Fock states, and the concept of an energy shell becomes meaningless: to exponential accuracy in $N$, remote Fock states, even if they are close in energy, have no common matrix elements with individual eigenstates.

The entanglement entropy then
 scales as $S_A \sim s(\delta/\delta_c) N_A/N$, 
 where $s$ is related to the entropy of the distribution of the localized eigenstate in Fock space. For $1\ll N_A\ll N$,  
$S_A \ll 1$ stays small down to $\delta\sim \delta_c$, where it jumps to 
$S_A \sim N_A$ at the localization transition to regime III.

\begin{figure}
\centering
\includegraphics[width=7.5cm]{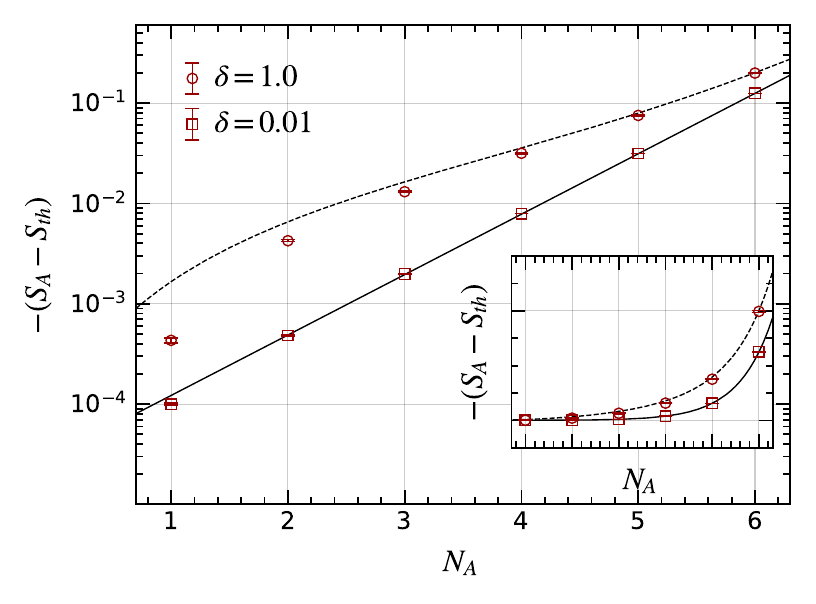}
\caption{\label{fig3}
Numerical entanglement entropies (symbols) vs.\ analytical (lines) for a system of size $N=15$
in regime I, $\delta=0.01$ (solid) and III, $\delta=1$ (dashed). Inset: linear scale representation of the same data.
}
\end{figure}

\emph{Numerical analysis: ---} Fig.~\eqref{fig3} shows a comparison of the analytical predictions of Eqs.~\eqref{S_page} and~\eqref{EERegimeIII} with numerical results obtained for the SYK Hamiltonian~\cite{SuppMat}. In that case, $\hat{H}_4
=
\frac{1}{4!}\sum_{i,j,k,l=1}^{2N} J_{ijkl} \hat{\chi}_i \hat{\chi}_j \hat{\chi}_k
\hat{\chi}_l $, where $\{\hat \chi_l\}$ are Majorana operators~\cite{SYK1,SYK2}. The competing
one-body operator reads $\hat{H}_2 = \sum_{i=1}^N m_i (2c^\dagger_ic^{\vphantom{
\dagger}}_i - 1)$,
where $c_i =
\tfrac{1}{2}(\hat{\chi}_{2i-1} + i\hat{\chi}_{2i})$ are complex fermion operators
defined by the Majoranas~\cite{SYK_GG,Shepelyansky17}. 
Referring to the supplemental material for details, the agreement is  
very good, and it becomes better with increasing $N_A$. (We have no certain explanation for the deviations at the smallest values of $N_A$.)

\emph{Discussion: ---} In this paper, we applied a combined analysis of the statistics and the entanglement properties of pure quantum states to explore the delocalized phase of disordered many body systems subject to long range correlations. Our analysis supports the view that the appealing concept of ``non-ergodic extended states'' --- adopted including in publications of the present authors~\cite{SYKRE,SYK4+2} --- should be abandoned in favor of a qualified interpretation of many body quantum ergodicity. Its key element is the support set $\{n\}$ of states of a given energy, the quantum analog of an energy shell. We have shown how the entanglement properties of pure quantum states   reveal ergodicity, and in addition characteristic correlations distinguishing the energy shells of genuine many body systems from those of phenomenological proxies.

What is the scope of the above findings? Referring to the supplemental material
for a more detailed discussion, the freedom to  adjust the exponents $\alpha,\beta$
entering the definition of the model Hamiltonian, implies that our result applies to
a wide class of \emph{effectively} long range interacting systems, among them
realizations whose interaction operators are short range in a microscopic (``real
space'') basis but long range in the eigenbasis of $\hat H_2$. It is tempting to
speculate on generalizations to yet wider system classes. To this end, we note that
the derivation of Eq.~\eqref{EERegimeIII} relies on a number of necessary conditions:
subsystem additivity $E\simeq E_A+E_B$ (requiring that the coupling energy between
the subsystems is negligibly small in the limit of large system sizes), statistically
independent distribution of the the energies $E_{A,B}$, and dependence of the
spectral density (measure) on no more than the single conserved quantity, energy.
Whether these criteria are not only required but actually sufficient to stabilize the result is
an interesting question left for forthcoming research~\cite{wip}. However, regardless of the
scope of Eq.~\eqref{EERegimeIII}, we reason that the combination of wave function
statistics and pure state entanglement defines the suitable diagnostic to
characterize the ergodic phase of many body quantum chaotic systems.

\emph{Acknowledgments: ---} D.~A.~H. thanks Vir Bulchandani and Sarang Gopalakrishnan   for   helpful
discussions. F.~M and T.~M.~acknowledge financial support by Brazilian agencies CNPq
and FAPERJ. A.~A. acknowledges partial support from the Deutsche
Forschungsgemeinschaft (DFG) within the CRC network TR 183 (project grant 277101999)
as part of projects A03. The work of M.~T. was supported in part by JSPS KAKENHI
Grant Numbers JP17K17822, JP20K03787, and JP20H05270.
D.A.H. is supported in part by DOE grant DE-SC0016244.

%% file: suppMat.tex
\title{Supplementary Material to:
``Quantum ergodicity in the many-body localization problem''
}

\date{\today}


\begin{abstract}
In this Supplemental Material, we provide details on the analytical calculation of the entanglement entropy 
and on the numerical calculations for the SYK model.
\end{abstract}

\maketitle

\section{Entanglement entropy and local density of states}
\label{app1}

In this section we discuss the derivation of Eq.~(2) describing the entanglement entropy in terms of 
the local density of states. The construction parallels a similar one for the 
wave functions of single particle systems~\cite{appEfetov,appMirlin}, 
(see also Ref.~\cite{App_SYK4+2} for a recent extension to Fock space), and we limit ourselves to an  outline of the main construction steps.

{\it Moments of the reduced density matrix from Fock space resolvents:---} Working in a first quantized representation --- where the Hamiltonian $\hat H$ is considered as a high dimensional matrix --- our starting point is a representation of the reduced density matrix,
$M_r(A)=\langle
{\rm Tr}_A( \rho_A^r)
\rangle$, 
in terms of retarded/advanced resolvent operators, 
$G_E^\pm=(E\pm i\eta -\hat H)^{-1}$. 
Introducing a formal Lehmann representation in term of exact eigenstates, it is straightforward to verify the likewise exact relation
\begin{align}
\label{app_eeGF}
M_r
&=
 {(2i)^{r-2}\over \pi \nu_E}
\lim_{\eta\to 0} \eta^{n-1}
\langle
{\rm tr}_A\left(
\left( 
{\rm tr}_B[ G_E^+ ]
\right)^{r-1}
{\rm tr}_B[ G^-_E ]
\right)
 \rangle,
 \end{align}
 where $\nu_E$ is the density of states at energy $E$.

\emph{Construction of the matrix integral:---} Following Efetov's supersymmetry approach~\cite{appEfetov,appMirlin}, 
we next represent the Green function matrix elements in Eq.~\eqref{app_eeGF} as Gaussian integrals. This representation is obtained from the   auxiliary formula 
 $M^{-1}_{nm}=
  \int D(\bar\psi,\psi)\,e^{-\bar \psi M\psi}\psi^\sigma_m\bar \psi^\sigma_n$, 
where $M$ is a general $L\times L$ matrix and the $2L$ dimensional `graded' vector $\psi=(\psi^\mathrm{b},\psi^\mathrm{f})^T$
contains $L$-commuting components $\psi_n^\mathrm{b}$, and an equal number of Grassmann 
components $\psi_n^\mathrm{f}$. 
The double integral over these variables cancels unwanted determinants 
$\det(M)$, while the pre-exponential factors, 
either commuting or anti-commuting, $\sigma=\mathrm{b,f}$, 
isolate the inverse matrix element. With the identification
 $M=\mathrm{diag}(-i [G^+]^{-1},i [G^-]^{-1})
 =
 \eta -i \sigma_3\hat H$, 
we are then led to consider the generating function
\begin{align}
\label{app_eq:Zpsi}
    {\cal Z}[j]
=
\int D(\bar{\psi},\psi)
\left\langle e^{-\bar{\psi}
\left(
i\eta\sigma_3
 -\hat H
\right)\psi + S_J}\right\rangle,
\end{align}
where we focus on the band center $E=0$, the
 average $\langle...\rangle$ is with respect to random coefficients of the Hamiltonian
 $\hat H$, $\sigma_3$ is a Pauli matrix distinguishing between advanced and retarded
 components, and $S_J=
\sum_n 
\left(
j_n \psi_n 
+
\bar{\psi}_n \bar{j}_n 
\right)$, 
with
 \begin{align*}
j_n
&=
(\alpha_n  \pi^{\rm rr}
+
 \beta_n \pi^{\rm aa})\otimes \pi^{\rm bb},\\
\bar{j}_n
&=
(\bar{\alpha}_n\pi^{\rm rr}  
+
\bar{\beta}_n\pi^{\rm aa})\otimes \pi^{\rm bb}.
\end{align*}
is a source term from which the required products are obtained by differentiation ($\partial_{\alpha_{n_i}}\equiv \partial_{\alpha_n}|_{n=n_i}$):
\begin{align*}
&\sum_{\sigma \in S_{r-1}}
G_E^R(n^1,m^{\sigma(1)})
...
G_E^R(n^{r-1},m^{\sigma(r-1)})
G_E^A(n^r,m^{r})
\nonumber\\
&\qquad =\prod_{l=1}^{r-1}\partial_{\bar{\alpha}_{m^l}}\partial_{\alpha_{n^l}}
\partial_{\bar{\beta}_{m^r}}\partial_{\beta_{n^r}}
{\cal Z}[j].
 \end{align*} 
Here, $\pi^{rr/aa}$ and $\pi^{bb}$ are projectors onto the subspaces of retarded/advanced and commuting variables, respectively. With this identity, and using the permutation symmetry of Green function matrix elements under trace, we obtain
\begin{align}
\label{app_ft_eeGF}
&M_r
=
c_r \lim_{\eta\to 0} \eta^{r-1}
\sum_{\{n^l\}} 
\prod_{l=1}^{r-1}\partial_{\bar{\alpha}_{m^l}}\partial_{\alpha_{n^l}}
\partial_{\bar{\beta}_{m^r}}\partial_{\beta_{n^r}}{\cal Z}[j],
 \end{align}
where  $c_r\equiv(2i)^{r-2}/(\pi \nu_E  (r-1)!)$ and  the differentiation  
arguments $m^i$ are  fixed as 
$m^l=(n^{l+1}_A,n^l_B)$, $l=1\ldots r-1$, and
$m^r=(n^1_A,n^r_B)$.
Using the notation in the man text this can be summarized as
$\mathcal{M}_B^i=\mathcal{N}_B^i$ and $\mathcal{M}_A^i=\mathcal{N}_A^{\tau i}$, where
$\tau i = (i+1)\mathrm{mod}(r)$.

{\it Effective action:---} We now average over the random parameters of the interaction Hamiltonian and then apply constructions steps standard in the theory of disordered electronic systems\cite{appEfetov,appMirlin}  and transferred to the SYK context in Refs.~\cite{App_AltlandBagrets,App_SYK4+2}. In regimes I-III, this procedure maps the generating function
onto the integral
${\cal Z}[j]=\int {\cal D}Q\, e^{- S[Q]+S_J[Q]}$, 
where $Q=Q\otimes 1_{\mathrm{Fock}}$ is a $4\times 4$ matrix in the spaces of advanced/retarded and commuting/anticommuting indices and  
\begin{align}
  \label{app_ErgodicFrequencyAction}
    S_\eta[Q] 
    = 
  \pi \eta\,{\rm STr}(\hat \nu Q\sigma_3),\quad
S_J[Q]
=
-i\pi{\rm STr}\left(
\bar{j} 
\hat{\nu} Q j
\right). 
\end{align}
Here `${\rm STr}$' refers to the graded trace over Fock and internal degrees of freedom,
and $\hat \nu$ is a diagonal matrix in Fock space with 
the local density of states as its
diagonal elements, 
$(\hat \nu)_n=\nu_n$.

{\it Moments:---}Performing  the $2r$-fold derivative, we arrive at
\begin{align}
\label{app_ft_moments}
&{\cal M}_r(A)
=
c_r
\lim_{\eta\to 0} \eta^{r-1}
\sum_{\sigma \in S_r}
\sum_{\cal N}
\nu_{n^1}...
\nu_{n^r}
\nonumber\\
\times&
\langle
Q^{\rm rr}_{{\rm bb}} ...
Q^{\rm rr}_{{\rm bb}} 
Q^{\rm aa}_{{\rm bb}}
\rangle
\delta_{{\cal N}_A, \sigma\circ \tau({\cal N}_A)}
\delta_{{\cal N}_B, \sigma({\cal N}_B)},
 \end{align}
 where the average $\langle...\rangle$ is over the action Eq.~\eqref{app_ErgodicFrequencyAction}. 
In a final step, we perform the matrix integral to obtain
 \begin{align}
c_r\eta^{r-1} 
&
\langle [Q^{\rm rr}_{\rm bb}]^{r-1}Q^{\rm aa}_{\rm bb}]\rangle
=
{1\over (D\nu)^r },
\end{align}
In this way, the identification of wave function moduli with coefficients of the local density of states fundamental to Eq.~(2)  of the main text is established.

\section{Entanglement entropy in regimes II/III}

{\it Leading contribution:---} Central to the analysis of the wave function moments is the reduced spectral density, $\lambda_{A,l}=\frac{1}{D\nu} \langle \delta_{\kappa}(v_l+v_B)\rangle_B$. Performing the Gaussian average Eq.~(5) of the main text, we obtain $D\nu=D\langle \delta_\kappa(v)\rangle_{AB}=\frac{D}{\sqrt{2\pi N}\delta}$ and similarly $\lambda_{A,l} =\frac{1}{D_A}\frac{\Delta}{\Delta_B}\exp(-v_l^2/2 \Delta_B^2)$. 

As discussed in the main text, the leading contribution to the entanglement entropy
comes from the identity permutation
\begin{align}
 M^{\rm id}_r(A)
&=
D_A D_B^r
\langle \lambda_A^r \rangle_A.
 \end{align}
Substituting the above result for $\lambda_A$ and performing the Gaussian average, we obtain 
\begin{align}
 \label{app_l}
 M^{\rm id}_r(A)
&=
\frac{D^{1-r}_A
 N_A^{r/2} }{ \sqrt{1+ rN_A/N_B}}.
\end{align}

{\it Subleading contribution:---} Single transpositions $\sigma=(ij)$, give the
subleading contribution to the entanglement entropy. Inspection of Eq.(2) of the main text (see also the index configuration defined by the right part of the bottom panel of Fig.1) shows that they provide a contribution  $M_r^\sigma=\sum_{l_1,l_2}\sum_{m_1,\ldots,m_{r-1}}\lambda_{l_1m_1}\nu_{l_2m_1} \lambda_{l_1m_2}\ldots\lambda_{l_1m_{r-1}}$ to the $r$th moment. Following the same recipe as above, we substitute $\lambda_{l,m}=(D\nu)^{-1}\delta(v_m+v_l)$ and the index summations by Gaussian averages over the energy variables $v_m\rightarrow v_A$ and $v_l\rightarrow v_B$. It is then straightforward to obtain 
\begin{align}
M^\sigma_r
&=
\frac{D_A^{2-r}}{D_B}
\frac{N^{r/2}}{\sqrt{ N_A}}
\frac{N_B^{(2-r)/2}}{\sqrt{2N_B + (r-1)N_A}}.
 \end{align}
 Noting that there are $\binom{r}{2}$ such terms, the differentiation in $r$ yields the entropy Eq.(6).

{\it Remaining contributions:---}In regime I, the leading and subleading
contributions discussed above give the Page entropy Eq.~(3) in the main
text~\cite{App_Page93}. Permutations that are not the identity or single
transpositions vanish. This cancellation has been discussed  in the string theory
literature~\cite{App_Stanford20,App_Liu20}, and the arguments presented there also
apply to regimes II $\&$ III. (Basically, the combinatorial factor for contributions
with a given number of transpositions are the Narayana numbers and vanish for more
than one transposition in the replica limit.) We thus conclude that Eq.~(6) describes
the entanglement entropy in regime III, at the same level of rigor as Page's
result in regime I.

{\it Comment on crossover to Regime I:---}The crossover between Page's result and our Eq.~(5) 
can be worked out, but requires a more elaborate analysis of above integrals without approximating the local density of 
 states by a $\delta$-function.  We leave this analysis for future work.

\section{Exact diagonalization}

\begin{figure*}
\centering
\includegraphics[width=18cm]{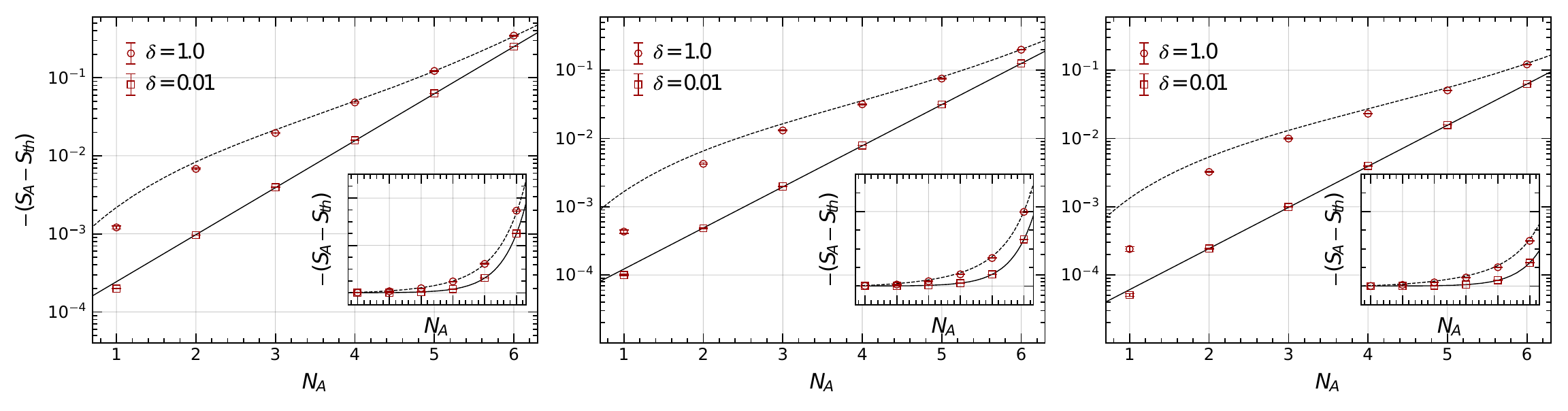}
\caption{\label{app_fig3}
Numerical entanglement entropies (symbols) vs.\ analytical (lines) for a system of size $N=14$ (left), $15$ (middle), $16$ (right) in regime I, $\delta=0.01$ (solid) and III, $\delta=1$ (dashed). Inset: linear scale representation of the same data.
}
\end{figure*}

We numerically calculated the reduced density matrix and the average entanglement
entropy for generic eigenstates (in the center of the band) of the SYK Hamiltonian $\hat{H}=\hat{H}_4+\hat{H}_2$, where
$\hat{H}_4
=
{1\over 4!}\sum_{i,j,k,l=1}^{2N}
J_{ijkl} \hat{\chi}_i \hat{\chi}_j \hat{\chi}_k \hat{\chi}_l$,
and the free particle contribution~\cite{App_SYK_GG,App_Shepelyansky17}
$\hat H_2 = \frac{1}{2}\sum_{i,j=1}^{2N}J_{ij}  \hat{\chi}_i  \hat{\chi}_j$.
Matrix elements $\{J_{ijkl} \}$ and $\{J_{ij} \}$ 
are drawn from Gaussian 
distributions with vanishing mean and
 variances $\langle |J_{ijkl}|^2 \rangle= 6J^2/(2N)^3$ and
 $\langle |J_{ij}|^2 \rangle=
\delta^2/2N$. 
The many body band width, $\Delta_4$, of the interaction operator and the
distribution width, $\Delta_2$, of the on-site random potential then read $\Delta_4 =
\sqrt{\frac{3J^{2}}{4N^3}\binom{2N}{4}}$ and $\Delta_2 =
\sqrt{\frac{\delta^{2}}{2N}\binom{2N}{2}}$, respectively. 
For our calculations, we
generate at least $100$ realizations of the Hamiltonians $(H_{4}, H_{2})$, 
taking the average of the entanglement entropy over eigenstates corresponding to energies within the middle $1/7$th of the
spectrum, unless otherwise mentioned. Here, the even and odd fermion parity sectors are diagonalized separately. 
We further improve the statistics by averaging over all $\binom{N}{N_{A}}$
Fock space bi-partitions. 

In Fig. 2 and Fig.~\ref{app_fig} (see below),
the error bar shows the standard deviation of the results over the realizations of the Hamiltonians.
The two parity sectors are treated as separate samples.
We observe an increasing ratio of this error bar to the value of $S_\mathrm{th}-S_A$ for
diminishing subsystem size $N_A$. The reason is the exponential
diminishing of $S_\mathrm{th}-S_A$ with decreasing $N_A$, which leads to relatively
larger numerical fluctuations around this value. 
We have no certain explanation for
the observation that in regime III (and I) results for smallest $N_A$
lie
outside the estimated error bar (see also Fig.~\ref{app_fig3}).

A subtlety in these calculations is that the SYK Hamiltonian conserves fermion parity. 
Considering the density matrix $\rho$ defined by  an eigenstate with definite parity, 
the partial trace 
leads to a block diagonal structure 
$\rho_A={\rm tr}_{B}\rho
=\left(\begin{smallmatrix} \rho_A^e &\\&\rho_A^o\end{smallmatrix}\right)$
with matrices $\rho_A^e$ and $\rho_A^o$ acting in the even and odd fermion parity 
subspaces of the subsystem $A$ Hilbert space. A trace over the two-dimensional parity sector
defines the  (normalized) reduced density matrix ${\rm tr}_P\rho_A = \rho_A^e + \rho_A^o$. 
One can then convince oneself that ${\rm tr}_P\rho_A$ has the same entropy as 
the reduced density matrix of a pure state in the $2^{N-1}$ system with broken fermion parity conservation.
This can be also verified by comparing our results in the fully ergodic
phase to Page's prediction for a Fock space of dimension $D=2^{N-1}$, as shown (by
the dashed line) in Fig.~2 of the main text.

{\it Variation of entanglement entropy with disorder:---} Our analytical analysis
predicts the formation of a $\delta$-independent plateau of the entanglement entropy
in regime III. In Fig.~\ref{app_fig} we show the numerically calculated entanglement
entropy for the system sizes $N=14, 15, 16$, as a function of $\delta$, and for
different partitions $N_A$. For the limited system sizes accessible to exact
diagonalization, the observation of a true plateau seems out of reach. However, one can see the formation of the plateau around
$\delta=1$ which becomes more pronounced with increasing $N_A$ and $N$. At the same time, the value $\delta=1$ defines the ``center'' of regime III. This follows from the recent work Ref.~\cite{App_SYK4+2} by some of the present
authors, where the regimes I-IV were characterized in terms of their WFMs. (On the same basis, $\delta=0.01$ 
is well within 
regime I.) While we cannot exclude a coincidence, the respective regime centers as determined by wave function statistics show the best agreement between numerics and analytics for the entropies.

\begin{figure}[t]
\centering
(a)\\
\includegraphics[width=8.5cm]{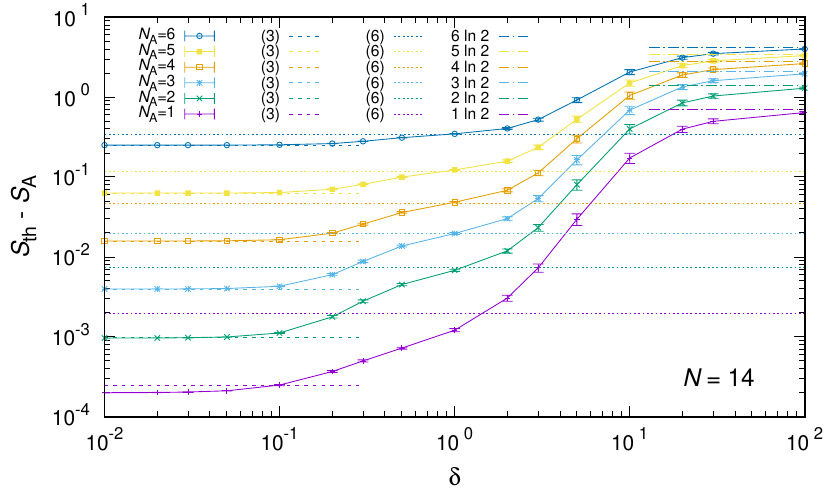}
(b)\\
\includegraphics[width=8.5cm]{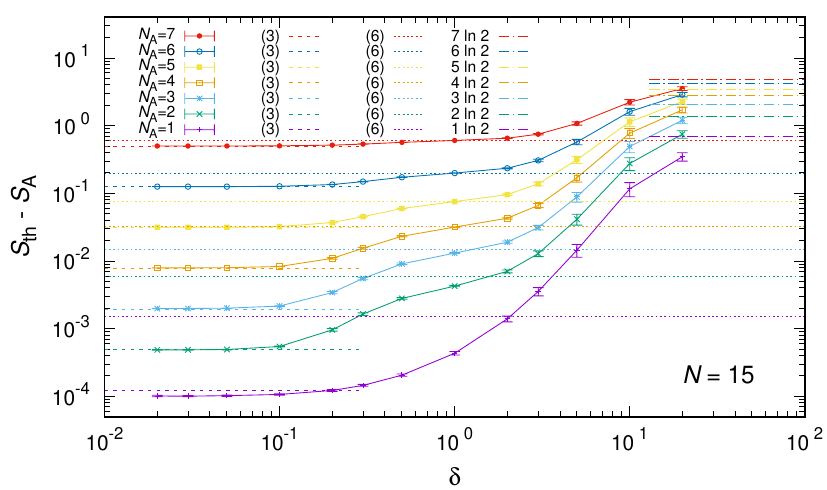}
(c)\\
\includegraphics[width=8.5cm]{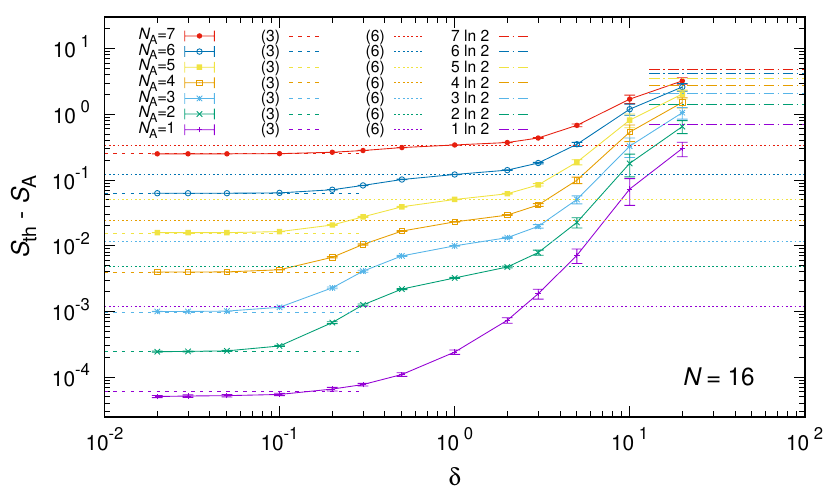}
\caption{\label{app_fig}
Entanglement entropy as a function of the disorder strength 
$\delta$ for different partitions $N_A$, 
(a) $N=14$, 104 realizations of the Hamiltonian and 1171 eigenvectors ($1/7$ of the entire spectrum) are used.
(b) $N=15$, 100 realizations of the Hamiltonian and 100 eigenvectors ($\sim 0.6 \%$) are used.
(c) $N=16$, at least 12 realizations of the Hamiltonian and 20 eigenvectors ($\sim 0.06\%$) are used.
}
\end{figure}

\section{Generalization} 
\label{sec:generalization}
 Within the above class of strong interaction coupled models, there is some freedom in the specific realization of the
$\hat H_2$ eigenstates. 
Broadly speaking, this setup is realized in 3 types of settings:  {\it i}) The system may not have any other geometry beyond that specified by the matrix elements of $\hat H_4$, as in a SYK model.  {\it ii})  The single-particle eigenstates of $\hat H_2$ may be localized in a $d$-dimensional real-space, and the couplings in $\hat H_4$ are such that the long-range couplings dominate (e.g., a power-law in space that decays slowly enough)~\cite{App_TM18}.  In these first two cases, as we take the limit of large $N$, the sparsity of the interactions can be adjusted with $N$ to set $\alpha$, but we require $\alpha>0$.  {\it iii}) The single-particle eigenstates of $\hat H_2$ may be all delocalized in real space. In this case, even local interactions couple all-to-all and for density-density interactions, for example, we will have $\alpha=4$.  In all three cases, as we take the limit of large $N$, the strength of the interactions can be adjusted with $N$ to set $\beta$. 
Our analysis does not apply to models of MBL with only short-range interactions~(see e.g. the recent numerical
study Ref.~\cite{App_Macet} on the multifractal scalings across the MBL transition). At the same time, it does not specifically exclude this case, and it seems natural that the ergodicity picture extends to it.
However the corroboration of that belief requires further study. 
(For very recent work on the entanglement entropy of extended random systems, see Ref.~\cite{App_HaqueKhaymovich2020}.)


%% file: merge.bbl
\begin{thebibliography}{99}
\bibitem{fn1}
Self consistency enters via the condition
 $\kappa_n=\pi \sum_m |J_{nm}|^2\nu_n$, where $|J_{nm}|$ are the matrix elements of the interaction operator (see also Ref.~\cite{SYK4+2}).
\end{thebibliography}
